\documentclass[useAMS,usenatbib]{mn2e}
\input psfig.sty

\usepackage{amssymb}
\usepackage{amsmath}
\usepackage{amsfonts}

\title[Cosmological Parameters and Cosmic Topology]
{Cosmological Parameters and Cosmic Topology}
\author[Rebou\c{c}as and Alcaniz]{M.J. Rebou\c{c}as$^{1}$\thanks{E-mail: reboucas@cbpf.br} and J. S. Alcaniz$^{2}$\thanks{E-mail: alcaniz@on.br}\\
$^{1}$Centro Brasileiro de Pesquisas F\'{\i}sicas, Rua Dr.\ Xavier Sigaud 150, 22290-180 Rio de Janeiro -- RJ, Brazil\\ $^{2}$Observat\'orio Nacional, Rua Gal. Jos\'e Cristino 77, 20921-400, S\~ao
Crist\'ov\~ao, Rio de Janeiro - RJ, Brazil}

\begin{document}

\date{Accepted ; Received}

\pagerange{\pageref{firstpage}--\pageref{lastpage}} \pubyear{2004}

\maketitle

\label{firstpage}

\begin{abstract}
Geometry constrains but does not dictate the topology of the $3$--dimensional
space. In a locally spatially homogeneous and isotropic universe, however, the 
topology of its spatial section dictates its geometry. We show that, besides 
determining the geometry, the  knowledge of the spatial topology through 
the circles--in--the--sky offers
an effective way of setting constraints on the density parameters associated 
with dark matter ($\Omega_m$) and dark energy ($\Omega_{\Lambda}$).
By assuming the Poincar\'e dodecahedral space as the circles--in--the--sky
detectable
topology of the spatial sections of the Universe, we re-analyze 
the constraints on the density parametric plane 
$\Omega_{m}$--$\,\,\Omega_{\Lambda}\,$ from the current type Ia 
supenovae (SNe Ia) plus X-ray gas mass fraction data, and show that
a circles--in--the sky detection of the dodecahedral space topology 
give rise to strong and complementary constraints on the region of the 
density parameter plane currently allowed by  these observational data sets.

\end{abstract}

\begin{keywords}
 Cosmology: theory - dark matter - distance scale
\end{keywords}

\section{Introduction}
The standard approach to cosmological modelling commences with the
assumption that our $3$--dimensional space is homogeneous
and isotropic at large scales. The most general spacetime metric 
consistent with the existence of a cosmic time $t$ and the principle 
of spatial homogeneity and isotropy is
\begin{equation}
\label{RWmetric} ds^2 = -dt^2 + a^2 (t) \left [ d \chi^2 +
f^2(\chi) (d\theta^2 + \sin^2 \theta  d\phi^2) \right ] \;,
\end{equation}
where $f(\chi)=(\chi\,$, $\sin\chi$, $\sinh\chi)$ depending on
the sign of the constant spatial curvature ($k=0,1,-1$), and $a(t)$
is the cosmological scale factor.  The metric~(\ref{RWmetric}) only 
express the above assumptions, it does not specify the underlying 
spacetime manifold $\mathcal{M}_4$ nor the corresponding spatial 
sections $M$. In this geometrical approach to model the physical 
world the dynamics of the Universe is clearly provided by a metrical 
theory of gravitation as, for example, General Relativity.

The Friedmann equation $k= H_0^2 a_0^2\, (\,\Omega_{\mathrm{tot}}-1\,)$%
\footnote{The total density at the present time $t_0$ is
$\Omega_{\mathrm{tot}}=\rho_{\mathrm{tot}}/\rho_{\mathrm{crit}}$ with
$\rho_{\mathrm{crit}} \equiv 3 H_0^2 / 8 \pi G$, and  $a_0$ and $H_0$
are, respectively, the scale factor and the Hubble parameter at $t_0$.}
makes apparent that the curvature (or the corresponding geometry)
of the spatial section $M$ of the Universe is an observable in
that for $\Omega_{\mathrm{tot}} > 1$  the spatial section is
positively curved, for $\Omega_{\mathrm{tot}} = 1$  it is flat
($k=0$), while for $\Omega_{\mathrm{tot}} < 1$ it is negatively
curved. Thus, a chief point in the search for the (spatial)
geometry of universe is to constrain the density $\Omega_{\mathrm{tot}}$ 
from observations. In the context of the current standard cosmological 
model, i.e., the $\Lambda$CDM scenario, this amounts to determining regions 
in the density parametric plane $\Omega_{m}\,$--$\,\,\Omega_{\Lambda}\,$,
which consistently account for the observations, and from which
one expects to infer the spatial geometry of the Universe.

In practice, to observationally  probe the spatial geometry of the Universe, 
by using, e.g., a single observational data set as the current type Ia 
supenovae (SNe Ia) observations, various degeneracies arise in the parametric 
plane $\Omega_{m}\,$--$\,\,\Omega_{\Lambda}\,$ (Riess et al. 2004). 
These degeneracies are  mitigated by either imposing reasonable priors, or 
combining the data with complementary observations, or both. An example in this 
regard is the combination of the current Supernovae Ia (SNIa)
compilation (Riess et al. 2004) with the Sloan Digital
Sky Survey (SDSS) galaxy power spectrum (Tegmark et al. 2004) and the cosmic
microwave background (CMB) measurements by the Wilkinson Microwave
Anisotropy Probe [WMAP] (Bennet et al. 2003; Spergel et al. 2003), which has
remarkably narrowed the bounds on the cosmological density
parameters $\Omega_m$ and $\Omega_{\Lambda}$. Besides being
sensitive to different combination of the density parameters,
these measurements probe the geometry of the Universe at
considerably different redshifts (typically $z < 2$ for SNe Ia or $\,z \sim 1100$ for CMB).

However, the spatial geometry constrains but does not dictate the topology
of the $3$--space $M$, and although the spatial section $M$ is usually
taken to be one of the simply-connected spaces, namely Euclidean
$\mathcal{R}^3$, spherical $\mathcal{S}^3$, or hyperbolic $\mathcal{H}^3$,
it is a mathematical result that great majority of locally homogeneous
and isotropic $3$--spaces $M$  are multiply-connected quotient manifolds
of the form $\mathcal{R}^3/\Gamma$, $\mathcal{S}^3/\Gamma$, and
$\mathcal{H}^3/\Gamma$, where $\Gamma$ is a fixed-point free group of
isometries of the corresponding covering space.
Thus, for example, for the Euclidean geometry ($k=0$) besides $\mathcal{R}^{3}$
there are 6 classes of topologically distinct compact orientable spaces $M$
that can be endowed with this geometry, while for both the spherical ($k=1$)
and hyperbolic ($k=-1$) geometries there are an infinite number of
topologically inequivalent manifolds with non-trivial topology that can be
endowed with each one of these geometries.

Now since the spatial geometry is an observable that constrains
but does not determine the topology of the $3$--space $M$, two
pertinent questions at this point are whether the topology may be
an observable and, if so, to what extent it can be used to remove
or at least reduce the degeneracies in the density parameter plane
$\Omega_{m}\,$--$\,\,\Omega_{\Lambda}\,$, which arise from
statistical analyses with data from current observations. 

The main aim of this paper, which  extends and complements our
previous work (Reboucas et al. 2005), is to address these questions in
the context of the $\Lambda$CDM scenario, by focusing our attention 
on the finite and positively curved Poincar\'e space model (Luminet et al. 2003) 
that accounts for the suppression of power at large scales observed by
WMAP(Bennet et al. 2003; Spergel et al. 2003), and also fits the WMAP temperature 
two-point correlation function (Aurich et al. 2005a; 2005b; Gundermann 2005). 
In other words, we shall show how to use the Poincar\'e dodecahedral 
space as a circles--in--the-sky observable topology of the Universe%
\footnote{Here, in line with the usage in the literature,
by topology of the universe we mean the topology of the space-like
section $M$.} 
to reduce the inherent degeneracies in the density parameters
$\Omega_m$ and $\Omega_{\Lambda}$ that arise from the so-called {\em gold} 
set of 157 SNe Ia, as compiled by Riess \emph{et al.\/} (2004), 
along with the latest Chandra measurements of the X-ray gas mass fraction 
in 26 galaxy clusters, as provided by Allen {\it et al.} (2004).

\section{Cosmic Topology}
\label{CosTop}

In a number of recent papers different strategies and methods to probe
a non-trivial topology of the spatial section of the Universe have been
discussed (see, e.g., Lehoucq et al, 1996; Roukema and Edge, 1997; Gomero et al., 2002; Fagundes and Gausmann, 1999; Uzan et al, 1999; Hajian and Souradeep, 2005; Hajian et al., 2005; and the review articles 
Lachi\`{e}ze-Rey and Luminet, 1995; Starkman, 1998; Levin, 2002; Rebou\c{c}as and Gomero, 2004).
An immediate observational consequence of a detectable non-trivial
topology%
\footnote{For a detailed discussion on the extent to which a non-trivial
topology may or may not be detected see Gomero et al. 2001a; 2001b; Weeks et al., 2003; Weeks, 2003.} 
of the $3$--space $M$ is that the sky will show multiple (topological)
images of either cosmic objects or repeated patterns of the cosmic
microwave background radiation (CMBR).
Here, we shall focus on the so-called ``circles-in-the-sky"  method,%
\footnote{For details on the other CMB 'approaches to cosmic topology' see,
e.g., Levin et al., 1998; Bond et al., 2000; Hajian and T. Souradeep, 2003; Oliveira-Costa et al. 1996; 2004; Dineen et al. 2004; Donoghue and Donoghue, 2004; Copi et al. 2003; Hipolito-Ricaldi and Gomero, 2005.}  
which relies on multiple copies of correlated circles in the CMBR
maps (Cornish et al. 1998), whose existence is clear from the following
reasoning: in a universe with a detectable non-trivial topology,
the sphere of last scattering necessarily intersects some of its
topological images along pairs of circles of equal radii, centered
at different points on the last scattering sphere (LSS), with the
same distribution of temperature fluctuations, $\delta T$, along
the circles correlated by an element $g$ of the covering group
$\Gamma$. Since the mapping from the last scattering surface to
the night sky sphere preserves circles (Penrose, 1959; Calv\~ao et al., 2005),
these pairs of matching circles will be written on the CMBR
anisotropy sky maps regardless of the background geometry and for
any non-trivial detectable topology. As a consequence, to
observationally probe a non-trivial topology on the available
largest scale, one should suitably scrutinize the full-sky CMB
maps in order to extract the correlated circles, whose angular
radii and relative position of their centers can be used to
determine the topology of the universe. Thus, a non-trivial
topology of the space section of the universe may be an observable,
which can be probed through the circles in the sky for all locally
homogeneous and isotropic universes with no assumption on the
cosmological density parameters.

Regarding the question as to whether the topology can be used as
an observable to reduce degeneracies in the cosmological density
parameters, we first note that the topology of a locally
homogeneous and isotropic $3$--manifold determines the sign of its
curvature (see, e.g., Bernshtein and Shvartsman, 1980), and
therefore the topology of the spatial section $M$ of the Universe
dictates its geometry. As a consequence, the detection of a
\emph{generic} cosmic topology alone would give rise to a strong
degeneracy on both density parameters, since it only determines
whether the density parameters take values in the regions below,
above, or on the flat line
$\Omega_{\mathrm{tot}}=\Omega_{m}+\Omega_{\Lambda}=1$. 

In what follows we examine to what extent a combination 
of the circles--in--the--sky detection of
a \emph{specific} spatial topology, namely, the Poincar\'e
dodecahedral space topology (which accounts for some observed
anomalies in WMAP CMB data) with the current SNIa and galaxy
clusters measurements may reduce the intrinsic density parameter
degeneracies of the $\Omega_{m}\,$--$\,\,\Omega_{\Lambda}\,$ plane.

\subsection{Poincar\'e Dodecahedral Space Model}

The Poincar\'e dodecahedral space $\mathcal{D}$ is a manifold of the
form $\mathcal{S}^3/\Gamma$ in which $\Gamma=I^\star$ is the binary
icosahedral group of order $120$. It is represented by a regular
spherical dodecahedron ($12$ pentagonal faces) along with the
identification of the opposite faces after a twist of $36^\circ$.
Such a space is positively curved, and tiles the $3$--sphere
$\mathcal{S}^3$ into $120$ identical spherical dodecahedra.

By assuming some priors and combining CMB data with other
astronomical data, the WMAP team reported (Spergel et al. 2003) both
the best fit value $\Omega_{\mathrm{tot}}= 1.02 \pm 0.02$
($1\sigma$ level), which includes a positively curved universe as
a realistic possibility, and account for the suppression of power at wide
angles ($\ell=2$ and $\ell=3$). These facts have motivated the
suggestion by Luminet \emph{et al.\/} (2003) of the
Poincar\'e dodecahedral space topology as an explanation for the apparent 
discrepancy between the $\Lambda$CDM  concordance model and WMAP
data. Since then, the dodecahedral space has been examined in some
works (Cornish et al, 2004; Roukema et al., 2004; Aurich et al, 2005a; 2005b; Gundermann, 2005), in which some further features of the model have
been carefully considered. As a result, it turns out that a
universe with the Poincar\'e dodecahedral space section accounts
for the suppression of power at large scales observed by WMAP, and
fits the WMAP temperature two-point correlation function for $
1.015 \leq\Omega_{\mathrm{tot}} \leq
1.020$ (Aurich et al, 2005a; 2005b), retaining the standard
Friedmann--Lema\^{\i}tre--Robertson--Walker
(FLRW) foundation for local physics.%
\footnote{A preliminary search failed to find the antipodal and nearly
antipodal matched circles with radii larger than $25^\circ$. This
absence of evidence of correlated may be due to several causes, among
which it is worth mentioning that Doppler and integrated Sachs-Wolfe
contributions to these circles may be strong enough to blur them,
and so the correlated circles can have been overlooked in the CMB maps
search (Aurich et al., 2005a). In this way, the `absence of evidence may not be evidence of absence',
specially given that effects such as Sunyaev-Zeldovich and
the finite thickness of the LSS, as well as possible systematics in the
removal of the foregrounds, can further damage the topological circle
matching.}
On this observational grounds, in what follows, we assume the 
Poincar\'e dodecahedron as the specific circles--in--the-sky
detected $3$--space topology of the Universe.

\section{Observations}

\subsection{Method}

\begin{figure}
\centerline{\psfig{figure=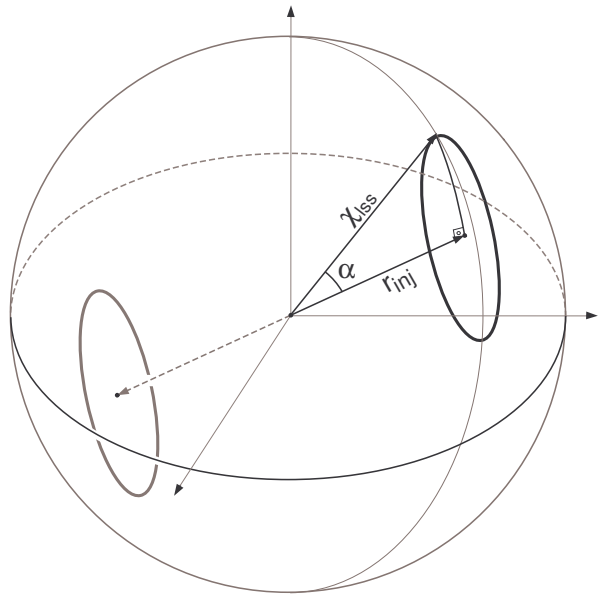,width=3.3truein,height=3.3truein,angle=0}
\hskip 0.1in} 
\caption{A schematic illustration of two antipodal
matching circles in the sphere of last scattering. These pair of circles come
about in all globally homogeneous positively curved manifolds with a 
detectable topology through the circles--in--the--sky method. The
relation between the angular radius $\alpha$ and the angular sides
$r_{inj}$ and $\chi^{}_{lss}$ is given by the following Napier's
rule for spherical triangles: $\cos \alpha = \tan
r_{inj} \cot \chi^{}_{lss}$}
\label{CinTheSky1}
\end{figure}

An important class of manifolds with a non-trivial topology is comprised
by the globally homogeneous manifolds. These manifolds satisfy a topological
principle of (global) homogeneity, in the sense that all points in $M$ are
topologically equivalent. In particular, in these spaces the pairs of matching
circles of the circles--in--the--sky method will be antipodal, as shown in
Figure~\ref{CinTheSky1}.

The Poincar\'e dodecahedral space (PDS) $\mathcal{D}$ is 
globally homogeneous,
and gives rise to six pairs of antipodal matched circles on the LSS, centered 
in a symmetrical pattern as the centers of the faces of the dodecahedron.
Figure~\ref{CinTheSky1} gives an illustration of two of these antipodal 
circles. Clearly the distance between the centers of each pair of circles
is twice the radius $r_{inj}$ of the sphere inscribable in $\mathcal{D}$.
Now, a straightforward use of a Napier's rule to the right-angled
spherical triangle shown in Fig.~\ref{CinTheSky1} furnishes a relation
between the angular radius $\alpha$ and the angular sides $r_{inj}$ and
radius $\chi^{}_{lss}$ of the last scattering sphere, namely
\begin{equation}
\label{Chialpha}
\chi^{}_{lss}=\tan^{-1} \left[\,\frac{\tan r_{inj}}{\cos \alpha}\,\right] \;,
\end{equation}
where $r_{inj}$ is a topological invariant, equal to  $\pi/10$ for
the dodecahedral topology, and the distance $\chi^{}_{lss}$ to the
origin \emph{in units of the curvature radius},
$a_0=a(t_0)=(\,H_0\sqrt{|1-\Omega_{\mathrm{tot}}|}\,)^{-1}\,$,
is given by
\begin{equation}
\label{ChiLSS}
\chi^{}_{lss}= \frac{d^{}_{lss}}{a_0} = \sqrt{|\Omega_k|}
\int_1^{1+z} \hspace{-4mm}
\frac{dx}{\sqrt{x^3 \Omega_{m} + x^2 \Omega_{k} +
 \Omega_{\Lambda}}} \;,
\end{equation}
where $d^{}_{lss}$ is the radius of the LSS, $x=1+z$ is an integration
variable, $\Omega_k = 1-\Omega_{\mathrm{tot}}$, and $z_{lss}=1089$ (Spergel et al. 2003).

Equations~(\ref{Chialpha}) and (\ref{ChiLSS}) give the relations
between the angular radius $\alpha$ and the cosmological density
parameters $\Omega_{\Lambda}$ and $\Omega_{m}$, and thus can be
used to set bounds on these parameters. To quantify this we
proceed in the following way. Firstly, we assume the angular
radius $\alpha = 50^\circ$, as estimated by Aurich et al (2005a).
Secondly, since the measurements of the radius $\alpha$
unavoidably involve observational uncertainties we take, in order
to obtain very conservative results, $\delta {\alpha} \simeq
6^\circ$, which is the scale below which the circles are
blurred (Aurich et al, 2005a).

\subsection{Statistical Analysis}

In order to study the effect of the PDS topology on the parametric 
space $\Omega_{m}\,$--$\,\,\Omega_{\Lambda}\,$, we use the  most recent 
compilation of SNe Ia data, the so-called \emph{gold} sample of 157 SNe Ia, 
recently published by Riess \emph{et al.} (2004) along with the
latest Chandra measurements of the X-ray gas mass fraction in 26 X-ray luminous, 
dynamically relaxed galaxy clusters spanning the redshift range $0.07 < z < 0.9$, 
as provided by Allen {\it et al.} (2004). We emphasize that this particular 
combination of observational data covers 
complementary aspects of the $\Omega_{m}\,$--$\,\,\Omega_{\Lambda}\,$ plane, 
in that while X-ray measurements are very effective to place limits on the 
clustered matter ($\Omega_m$) the new SNe Ia sample tightly constrains the 
unclustered component ($\Omega_{\Lambda}$).

\begin{figure*}
\centerline{\psfig{figure=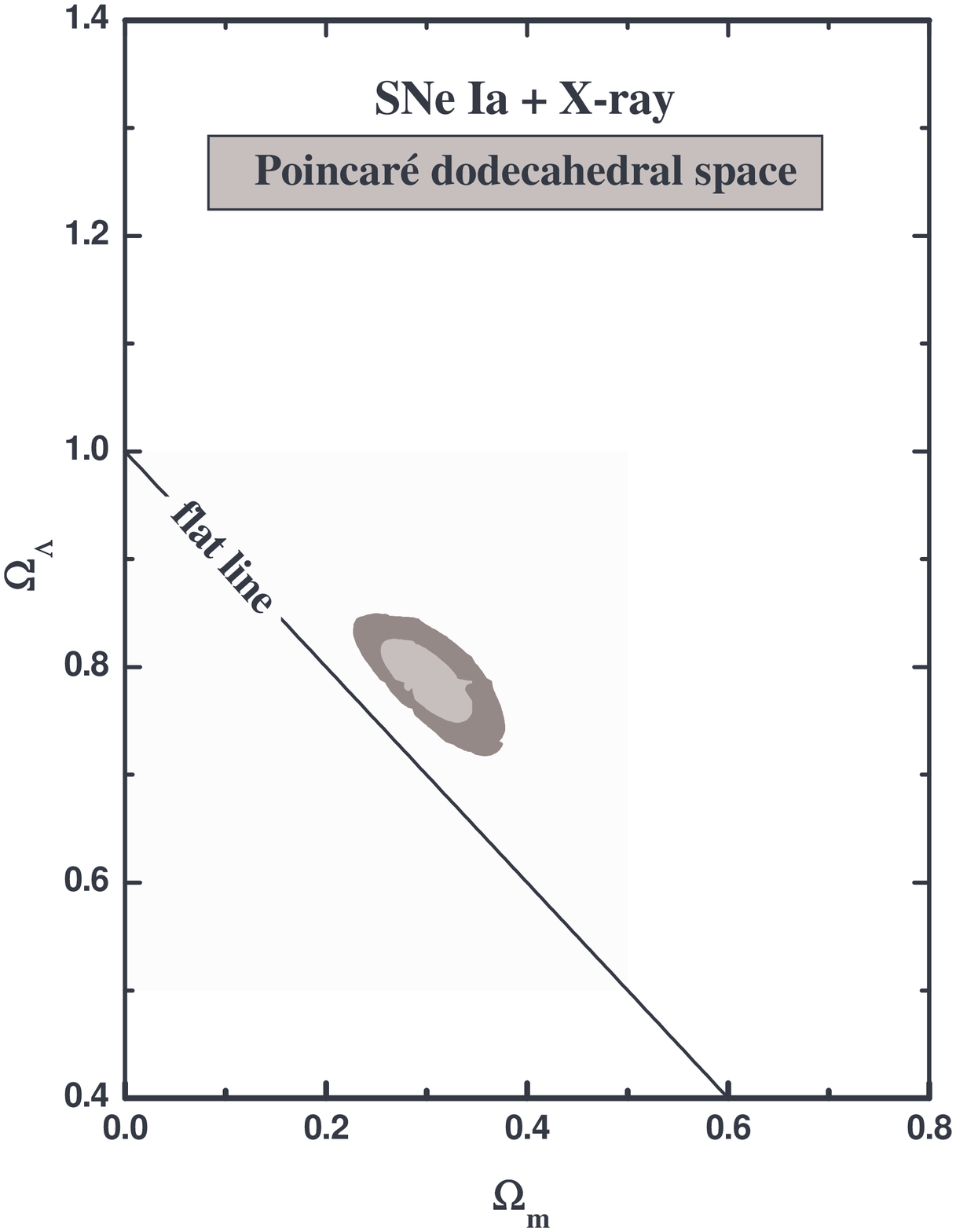,width=3.2truein,height=3.4truein,angle=0}
\psfig{figure=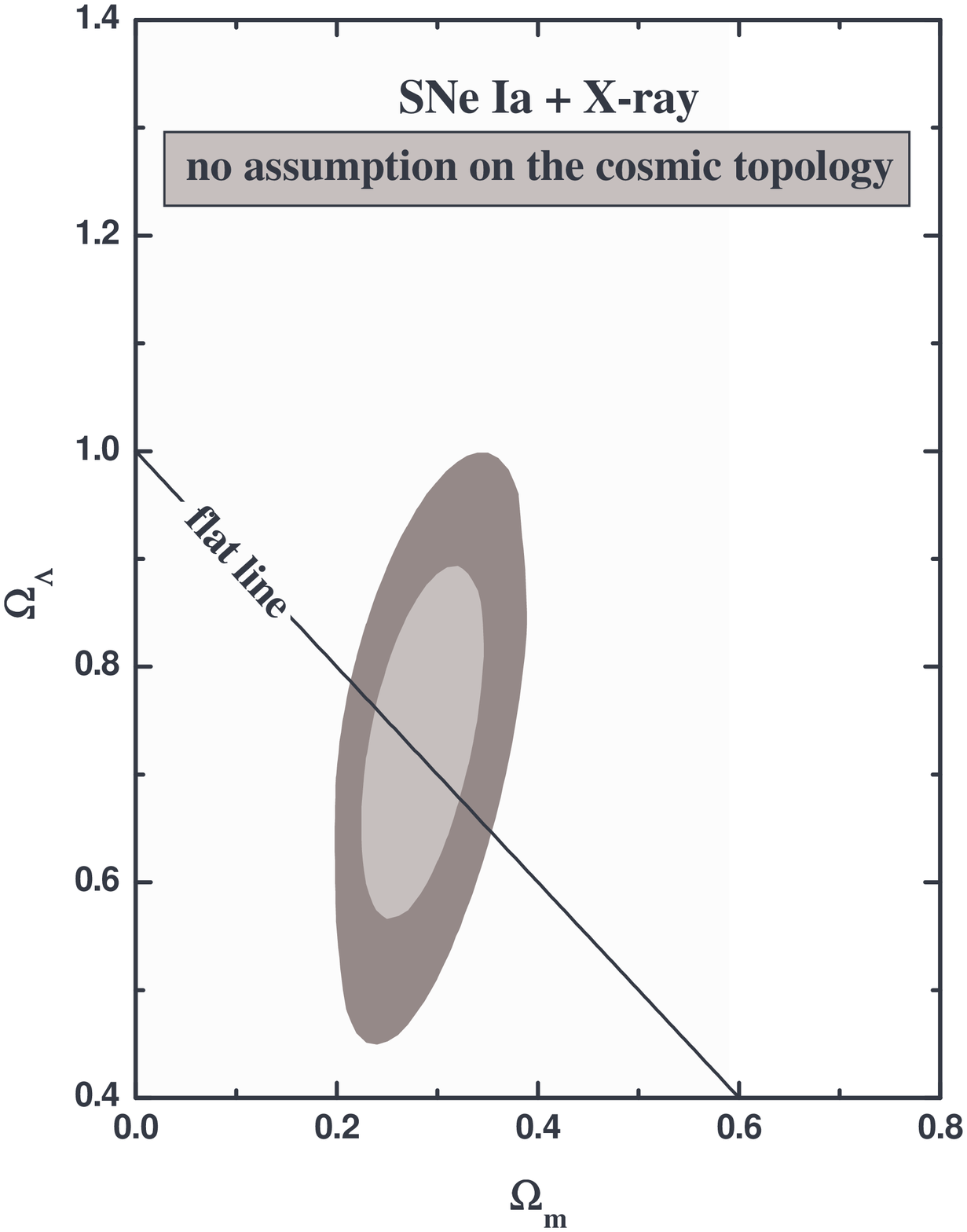,width=3.2truein,height=3.4truein,angle=0}
\hskip 0.001in} 
\caption{The results of our statistical analyses. The panels show confidence 
regions ($68.3\%$ and $95.4\%$ c.l.) in the $\Omega_{m}\,$--$\,\,\Omega_{\Lambda}\,$ 
plane from the latest Chandra measurements of the X-ray gas mass fraction in 26 
galaxy clusters ($0.07 < z < 0.9$) plus determinations of the baryon density 
parameter, measurements of the Hubble parameter and the \emph{gold} sample of 157 
SNe Ia. In the left panel a specific circles--in--the--sky detection of the 
dodecahedral space topology  with $\alpha=50^\circ$, $\delta \alpha= 6^\circ$
is assumed. 
The best fit values for the matter and vacuum density parameters are, respectively, 
$\Omega_m = 0.30 \pm 0.04$ and $\Omega_{\Lambda} = 0.79 \pm 0.03$ at 95.4\% 
c.l., which provides $\Omega_{\rm{tot}} \simeq 1.09 \pm 0.05$. In the right panel 
the conventional SNe Ia plus X-ray analysis is shown for comparison.}
\label{statisticalAnalysis}
\end{figure*}

\subsection{$f_{\rm gas}$ versus redshift test}

The X-ray gas mass fraction test $[f_{\rm gas}(z)]$ was first introduced by Sasaki (1996) 
and further
developed by Allen et al. (2002a;2002b) (see also Ettori et al. 2003; Lima et al. 2003; 2004; Chen and Ratra 2004;  Zhu and Alcaniz 2005; Alcaniz and Zhu 2005). This is based on the
assumption that rich clusters of galaxies are large enough to provide a fair
representation of the baryon and dark matter distributions in the Universe (Fukugita et al. 1998). Following
this assumption, the matter content of the Universe can be expressed as
the ratio between the baryonic content and the gas mass fraction, i.e., $\Omega_{\rm m}
\propto  \Omega_{\rm b}/f_{\rm gas}$. Moreover, as shown by Sasaki 1996, since $f_{\rm gas} \propto
d_{\rm{A}}^{3/2}$ the model function can be defined as (Allen \emph{et al.} 2002a) 
\begin{eqnarray} \label{fgas}
f_{\rm gas}^{\rm mod}(z) = \frac{ b\Omega_{\rm b}} {\left(1+0.19
\sqrt{h}\right) \Omega_{\rm m}} \left[ 2h \,
\frac{d_{\rm A}^{\rm{E-dS}}(z)}{d_{\rm A}^{\rm{\Lambda CDM}}(z)} \right]^{1.5},
\end{eqnarray}
where the bias factor $b$ is a parameter motivated by gas dynamical simulations that
takes into account the fact that the baryon fraction in clusters is slightly depressed
with respect to the Universe as a whole (Eke \emph{et al.} 1998; Bialek \emph{et al.} 
2003), the term $0.19\sqrt{h}$ stands
for the optically luminous galaxy mass in the cluster and the ratio between the angular diameter distances ${d_{\rm A}^{\rm{E-dS}}(z_{\rm i})}/{d_{\rm A}^{\rm{\Lambda CDM}}(z_{\rm i})}$ accounts for deviations in the geometry of the Universe (here modelled by the $\Lambda$CDM model) from the default
cosmology used in the observations, i.e., the  Einstein-de Sitter (E-dS) model (see 
Allen \emph{et al.} (2004)  for more observational details).

In order to perform the $f_{\rm gas}$ test, three Gaussian priors are added to our analysis, namely,  on the baryon density parameter, $\Omega_bh^{2} = 0.0224 \pm 0.0009$ (Spergel 
\emph{et al.}  2003), on the Hubble parameter, $h = 0.72 \pm 0.08$ Freedman \emph{et al.} (2001), and on the bias factor, $b = 0.824 \pm 0.089$ (Eke \emph{et al.} 1998; Bialek 
\emph{et al.}  2003). Thus, the total minimization $\chi^2_{f_{\rm gas}}$ is written as
\begin{eqnarray}
\chi^2_{f_{\rm gas}} & = &\sum_{i = 1}^{26}
\frac{\left[f_{\rm gas}^{\rm mod}(z_{\rm i})- f_{\rm gas,\,i}
\right]^2}{\sigma_{f_{\rm gas,\,i}}^2} 
+ \left[\frac{\Omega_{b}h^{2} - 0.0224}{0.0009}\right]^{2} 
+ \nonumber \\ & & 
\quad  \quad \quad + \left[\frac{h - 0.72}{0.08}\right]^{2} + \left[\frac{b -
0.824}{0.089}\right]^{2},
\end{eqnarray}
where $f_{\rm gas}^{\rm mod}(z_{\rm i})$ is given by Eq. (\ref{fgas}) and $f_{\rm gas,\,i}$ is
the observed values of the X-ray gas mass fraction with errors $\sigma_{f_{\rm
gas,\,i}}$.

\subsection{Magnitude versus redshift test}

Supernovae observations provide the most direct evidence for the current cosmic acceleration. To perform a statistical analysis with the current supernovae data we first define the predicted distance modulus for a supernova at redshift $z$, given a set of
parameters $\mathbf{s}$, i.e.,
\begin{equation} \label{sne}
\mu_p(z|\mathbf{s}) = m - M = 5\mbox{log} d_L + 25\;,
\end{equation}
where $m$ and $M$ are, respectively, the apparent and absolute magnitudes, the complete
set of parameters is $\mathbf{s} \equiv (h, \Omega_{m}, \Omega_{\Lambda})$ and $d_L$ stands for the
luminosity distance (in units of megaparsecs). 

The set of parameters $\mathbf{s}$ is estimated by using a $\chi^{2}$ statistics, with
\begin{equation}
\chi_{SNe}^{2} = \sum_{i=1}^{157}{\frac{\left[\mu_p^{i}(z|\mathbf{s}) -
\mu_o^{i}(z|\mathbf{s})\right]^{2}}{\sigma_i^{2}}}\;,
\end{equation}
where $\mu_p^{i}(z|\mathbf{s})$ is given by Eq. (\ref{sne}), $\mu_o^{i}(z|\mathbf{s})$ is the
extinction corrected distance modulus for a given SNe Ia at $z_i$, and $\sigma_i$ is
the uncertainty in the individual distance moduli, which includes uncertainties in
galaxy redshift due to a peculiar velocity of 400 km/s (For more details on statistical analyses involving SNe Ia observations we refer the reader to Padmanabhan and Choudhury, 2003; Zhu and Fujimoto, 2003; Nesseris and Perivolaropoulos, 2004; Alcaniz, 2004; Alcaniz and Pires, 2004; Choudhury and Padmanabhan, 2005; Shafieloo et al., 2005).

\subsection{Topological constraint}

Similarly to the Gaussian priors added to the $f_{gas}$ test, the Poincar\'e dodecahedral space topology is included in our statistical analysis as a prior relative to the value of $\chi^{}_{lss}$,
which can be easily obtained from a elementary combination of Eqs.~(\ref{Chialpha})~--~(\ref{ChiLSS}). In other words, the contribution of the topology to our statistical analysis is a term of the form
\begin{equation}
\chi^2_{\mathrm{topology}} = \frac{(\chi^{\mathrm{Th}}_{lss} - \chi^{\mathrm{Obs}}_{lss})^2}{(\delta \chi_{lss})^2}\;,
\end{equation}
where $\chi^{\mathrm{Th}}_{lss}$ is given by Eq.~(\ref{Chialpha})
and the uncertainty $\delta \chi_{lss}$ comes from the uncertainty $\delta \alpha$
of the circles--in--the--sky. This means that the total $\chi_{total}^2$ minimization function is given by
\begin{equation}
\chi_{total}^2 = \chi^2_{f_{\rm gas}} + \chi^2_{SNe} + \chi^2_{\mathrm{topology}}.
\end{equation}

\section{Results and Discussions}

The left panel in Figure~\ref{statisticalAnalysis} shows the results of our 
statistical analysis. Confidence regions -- 68.3\% and 95.4\% confidence 
limits (c.l) -- in the parametric space $\Omega_{m}\,$--$\,\,\Omega_{\Lambda}\,$
are displayed for the particular combination of observational data described 
above. For the sake of comparison, we also show in the right panel the $\Omega_{m}\,$--$\,\,\Omega_{\Lambda}\,$ plane for the conventional SNe Ia plus 
Galaxy Clusters analaysis, i.e., the one without the above cosmic topology
assumption. By comparing both analyses, it is clear that our initial premiss 
that a circles--in--the--sky detection of a non-trivial space topology reduces 
considerably the parametric space 
region allowed by the current observational data, and also breaks some 
degeneracies arising from the current SNe Ia and X-ray gas mass fraction 
measurements. The best-fit parameters for this SNe Ia+X-ray+Topology analysis 
are $\Omega_m = 0.30$ and $\Omega_{\Lambda} = 0.79$ with the reduced 
$\chi_{\nu}^2 \equiv \chi_{min}^2/\nu \simeq 1.12$ ($\nu$ is defined 
as degree of freedom). Note that such a value of $\chi_{\nu}^2$ is 
slighly smaller than the one obtained by fitting the \emph{gold} sample 
of SNe Ia to the flat $\Lambda$CDM (concordance) scenario, i.e., 
$\chi_{\nu}^2 \simeq 1.14$, and equal to the value found for the 
$\Lambda$CDM model with arbitrary curvature (Riess et al, 2004). 
At $95.4\%$ c.l. we also obtain $0.26 \leq \Omega_m \leq 0.34$ and 
$0.76 \leq\Omega_{\Lambda} \leq 0.82$ providing $\Omega_{\rm{tot}} 
\simeq 1.09 \pm 0.05$, which is consistent at 2$\sigma$ level with 
the value reported by the WMAP team, i.e., $\Omega_{\mathrm{tot}}=
1.02 \pm\, 0.02$ at 1$\sigma$ (Spergel et al 2003). Note also that the 
above best-fit scenario is in full agreement with most of the current 
observational analyses (see, e.g., Roos (2005) for an updated review) 
and corresponds to a current accelerated  universe with $q_o \simeq -0.64$ 
and a total expanding age of $9.67h^{-1}$ Gyr. Finally, we also note 
that for values of the angular radius in the range $0^\circ \leq  \alpha \leq 90^\circ$, Eq.(\ref{Chialpha}) (for, e.g.,  $\Omega_m=0.28$) gives the interval $1.01 \lesssim \Omega_{\mathrm{tot}} \lesssim 1.3$, 
making clear that the very detection of first 6 pairs of circles 
predicted by the Poincar\'e dodecahedral space topology would 
restrict the allowed range for the total density parameter. This 
sensitivity of $\alpha$ with the value of $\Omega_{\mathrm{tot}}$ was recently discussed by Roukema et al. (2004), and reinforced by  
Aurich et al (2005a). Clearly,  additional 
limits on the total density arise when one assumes
a specific value of $\alpha$ and a related uncertainty 
$\delta \alpha$.

\section{Concluding Remarks}

By assuming the Poincar\'e dodecahedral space as circles--in--the-sky detected
topology of the spatial sections of the Universe, we have re-analyzed 
the constraints on the parametric space $\Omega_{m}\,$--$\,\,\Omega_{\Lambda}\,$ 
from the current SNe Ia and X-ray gas mass fraction data. As the main outcome 
of this analysis, we have shown that once detected the PDS topology along with the corresponding circles--in--the-sky, they 
give rise to very strong and complementary constraints on the region of density 
parameter plane currently allowed by the cosmological observations. 
We have also discussed how the degeneracies inherent to a joint analysis involving 
SNe Ia observations and X-ray gas mass fraction measurements are drastically 
reduced by the assumption of a circles--in--the-sky detection of a PDS 
topology. According to Eq.(\ref{Chialpha}), the main reason for this 
breakdown of the degeneracies is the sensitivity of the angular radius with the 
total density parameter. Finally, our results also indicate 
that cosmic topology may offer a fruitful strategy to constrain the 
density parameters associated with dark energy and dark matter.

\section*{Acknowledgments}
The authors thank CNPq for the grants under which this work was carried out. JSA is also supported by Funda\c{c}\~ao de Amparo \`a Pesquisa do Estado do Rio de Janeiro (FAPERJ). The authors are also grateful to A.F.F. Teixeira for valuable discussions.

\end{document}